# Investigation of Martian UV Dayglow Emissions in the Southern Hemisphere during Solar Quiet-time Conditions: Insights from Multi-year MAVEN/IUVS Observations

Aadarsh Raj Sharma[1], Lot Ram[1], Sumanta Sarkhel[1,2*]

*Sumanta Sarkhel, Department of Physics, Indian Institute of Technology Roorkee, Roorkee - 247667, Uttarakhand, India (sarkhel@ph.iitr.ac.in)

[1]Department of Physics, Indian Institute of Technology Roorkee, Roorkee, Uttarakhand, India

[2]Centre for Space Science and Technology, Indian Institute of Technology Roorkee, Roorkee, Uttarakhand, India

**Abstract**

The southern hemisphere of Mars possesses concentrated region of strong crustal magnetic fields (CMF), which generate localized magnetic anomalies that can influence atmospheric dynamics and energy deposition in the Martian thermospheric-ionospheric system. Although their effects on the atmosphere (>200 km) in the southern hemisphere are well documented, however their role in modulating the behavior of atmospheric plasma and neutrals below 200 km are poorly understood. The atmosphere at these altitudes can be comprehended by studying the variation of dayglow emissions. We have investigated few dayglow emissions over the CMF and non-CMF regions using the MAVEN remote-sensing measurements from Martian Years 33-37. Particularly, the CO Cameron bands, $CO_2^+$ ultraviolet doublet, and atomic oxygen emissions at 297.2 nm, 130.4 nm, and 135.6 nm have been studied below 200 km during solar quiet-time conditions. The results show strong seasonal variations in all the emissions peak altitudes and intensities in the dayside and near-terminator regions. The variation in the peak altitude of molecular and atomic emissions are nearly 20 km and 30 km, respectively. On the dayside, the emissions show minimal variation across CMF and non-CMF regions, indicating a minimal effect of the CMF at these altitudes, suggesting a key role of plasma demagnetization and photochemical processes. In addition, an insignificant variation in the dayglow emissions is likely masked by compensating mechanisms such as energy-dependent electron shielding and thermospheric expansion. This work presents the first focused investigation on the dayglow emissions over CMF and non-CMF regions across different seasons and solar zenith angles.

**Key points**

1. Seasonal variations of Martian dayglow have been studied over crustal and non-crustal magnetic field regions from Martian Years 33-37.
2. Insignificant variations were observed in dayglow over crustal and non-crustal field regions with variation of peak across seasons.
3. Molecular and atomic emissions show a deviation of ~60-70% and ~30-40%, respectively, from the dayside to near-terminator region.

**Plain Language Summary**

Mars lacks a global magnetic field like Earth. However, its crust contains strong localized magnetic fields, especially in the southern hemisphere. These crustal magnetic fields (CMF) can create mini-magnetospheres that influence the Martian atmosphere, particularly at high altitudes. Although their effects above 200 km are well studied, their impact on atmosphere below 200 km remains unclear. In this study, we analyzed five Martian years of UV data from NASA's MAVEN spacecraft to investigate how CMFs affect five key emissions from atoms and molecules below 200 km. We found that, on the dayside, emission peak intensities and altitudes vary mainly with season and solar input, not with magnetic field strength, suggesting CMF have minimal direct impact due to the low magnetic control of plasma in this region. However, near the dayside-terminator region, dayglow emissions respond more strongly to local magnetic field conditions. These responses vary by season and region, indicating that CMF can influence how energy from space particles is deposited in the atmosphere, likely by affecting electron precipitation and local atmospheric composition. Even so, these effects are complex and often balanced by other atmospheric processes, limiting their overall impact.

# 1. Introduction

The southern hemisphere of Mars hosts strong localized crustal magnetic anomalies, concentrated between 25°S–90°S and 120°E–240°E (Brain et al., 2003; Cain et al., 2003). These magnetic fields, first detected by Mars Global Surveyor (MGS; Acuña et al., 1998, 1999), were fully mapped during its >7-year mission at ~400 km altitude (Connerney et al., 2005). Unlike a global magnetic field, these anomalies are highly localized and exert significant influence on the Martian atmosphere and space environment. Although weaker than Earth's global magnetic field, Martian crustal magnetic fields (CMF) exert a strong influence on the plasma dynamics in the upper atmosphere and affect the structure of the ionosphere, induced magnetosphere, boundary regions, and interactions with the solar wind (Opgenoorth et al., 2010; Brain et al., 2003; Crider et al., 2002; Edberg et al., 2008, 2009; Garnier et al., 2022; Hall et al., 2016; Ma et al., 2014).

Above 200 km, electron density enhancements have been observed in crustal field regions using MARSIS (Andrews et al., 2013, 2015; Dubinin et al., 2012) and MAVEN/LPW (Andrews et al., 2015; Flynn et al., 2017), with concurrent reductions in electron temperature (Flynn et al., 2017; Sakai et al., 2019; Andrews et al., 2023; Nayak et al., 2024). Withers et al. (2019) also observed elevated ion densities ($O^+$, $CO_2^+$, $O_2^+$) using NGIMS. These variations are significant above 200 km where transport processes dominate; below this altitude, CMF have minimal impact on the electron density and temperature, suggesting limited influence on dayside photochemistry.

However, some cusp-like magnetic field structures may elevate electron densities near the ionospheric peak (Nielsen et al., 2007; Gurnett et al., 2008). Chen et al. (2024) showed that CMF intensity and configuration can modulate peak density by altering solar wind energetic particle precipitation, with strong horizontal fields reducing impact ionization. Apart from the studies focused on showing the impact of crustal magnetic field on plasma densities and temperature, various studies have elucidated over the relation of neutral densities and temperature with crustal field. Flynn et al., (2017), found that neutral densities and temperatures do not depend on the magnetic field conditions. The study on the possible impact of crustal fields on the thermal structure of dayside Martian upper atmosphere was performed using the dayglow emission data of MEX/SPICAM (Leblanc et al., 2006; Stiepen et al., 2015). Leblanc et al., (2006) found strong dependence between high neutral temperature and crustal magnetic field. On the contrary, a lack of any clear correlation between neutral temperature and crustal magnetic field was reported by Stiepen et al., (2015). Cui et al., (2018) using the deep dip (DD) campaign of MAVEN/NGIMS found that the modulation of neutral temperature by crustal magnetic fields occurs with limited altitude range is around 20-40 K in dayside and terminator region.

Although, the influence of CMF above 200 km is well established, their effects on the plasma and neutral dynamics below this altitude remain uncertain. While CMF-linked auroral emissions are well known on the nightside (Bertaux et al., 2005; Soret et al., 2016; Chirakkil et al., 2024), their effect on dayglow emissions is less understood (Leblanc et al., 2006; Stiepen et al., 2015). Since dayglow is directly driven by plasma and neutral interactions (Leblanc et al., 2006; Gérard et al., 2019; Gkouvelis et al., 2018; Ritter et al., 2019), it offers a valuable tool for probing CMF-related variability.

This study takes advantage of large dataset of Imaging Ultraviolet Spectrograph (IUVS; McClintock et al., 2015) onboard Mars Atmosphere and Volatile EvolutioN (MAVEN; Jakosky, Lin, et al., 2015) and provides comprehensive analyses over the impact of crustal magnetic field on prominent dayglow emissions of Mars in the southern hemisphere during solar quiet-time. In this paper, we analyze the possible impact of CMF on various dayglow emissions over different Martian seasons, solar zenith angle (SZA) and longitudinal ranges. This work will further provide possible mechanisms responsible for the dayglow variability over the crustal magnetic field region.

## 2. Data and Methodology

Since its arrival in orbit around Mars in 2014, MAVEN is performing an elliptical orbit with a period of 4.5 hours (currently ~3.5 hours), an apoapsis altitude of approximately 6,200 km, a periapsis altitude around 150 km (currently ~180 km), and an orbital inclination of 75°. The data utilized in this study were acquired from the remote-sensing measurements of IUVS instrument onboard the MAVEN spacecraft. The MAVEN/IUVS instrument is designed to observe the Martian upper atmosphere by capturing far ultraviolet (FUV, 115-190 nm) and middle ultraviolet (MUV, 190-340 nm) spectra. The instrument is capable of operating in multiple observational modes, including limb scans, echelle mode, disk scans, coronal scans, and stellar occultations. For the purposes of this study, only the limb scan mode datasets are being considered.

In this study, we use the limb daytime radiance profiles of key atmospheric emissions from Mars, specifically focusing on the primary emission lines of oxygen and carbon dioxide, namely the CO Cameron band, $CO_2^+$ UVD, $O(^1S)$ 297.2 nm, $O(^3S)$ 130.4 nm, and $O(^5S)$ 135.6 nm, within the altitude range of 60–200 km. The five Martian years (MY 33-37) data have been considered in the present study, that covers the southern hemisphere of Mars, majorly from 25°S to 90°S latitudes.

In order to maintain the spatial variability of the data across the southern hemisphere, we further divided it into three distinct longitudinal (ϕ) bands. The first longitude band (120° < ϕ < 240°) corresponds to regions lie over the strong crustal magnetic field, while the other two bands (longitudes: ϕ < 120° and ϕ > 240°) represent the non-crustal magnetic field (non-CMF) regions (Brain et al., 2003; Connerney et al., 2001; Nayak et al., 2024). The longitudinal categorization is carefully designed to ensure a uniform data distribution across three considered longitudinal bands, that will eventually minimize possible bias arising from uneven data distribution.

Apart from the longitudinal categorization, we have further divided the data based on the solar zenith angle (SZA), to ensure the distinct measurements of the dayside (SZA < 70°) and dayside-terminator regions with SZA range of 70°-90° (Krymskii et al., 2003; Benna et al., 2015; Duru et al., 2019). To facilitate a more structured and seasonal analysis of the limb radiance profiles, we have categorized the data into four distinct Martian seasons based on the solar longitude (Ls): Southern Autumn (Ls: 0°–90°), Southern Winter (Ls: 90°–180°), Southern Spring (Ls: 180°–270°), and Southern Summer (Ls: 270°–360°). Finally, after going through aforementioned selection, we have obtained the median radiance profile. Given the larger dataset used in this study, a median radiance profile is less affected by outliers and better captures the typical pattern of dayglow brightness. Additionally, past studies have reported the impact of space weather events and global dust storm on Martian dayglow and atmosphere (Jain et al., 2018; Gkouvelis et al., 2020; Ram et al., 2024a; Sharma et al., 2025a; Lou et al., 2025). Therefore, to ensure that these profiles reflect undisturbed atmospheric conditions, we have selected only quiet-time IUVS observations from MAVEN, excluding orbits affected by space weather events such as solar flares, coronal mass ejections (CME), as well as Martian global dust storms (GDS).

## 3. Results

In order to investigate the effect of crustal magnetic field on $O(^3S)$ 130.4 nm, $O(^5S)$ 135.6 nm, $CO_2^+$ UVD, the CO Cameron band, and $O(^1S)$ 297.2 nm dayglow emissions, we have distributed the data as described in Section 2. The following section presents the variation of dayglow emission in the southern hemisphere between crustal and non-crustal longitudes distributed over four Martian seasons. Figure 1 shows the data distribution in CMF and non-CMF (ϕ < 120°; ϕ > 240°) regions, in four Martian seasons, along with dayside (SZA < 70°) and terminator region (SZA: 70°-90°). The bars plots in red, blue and purple show the longitude bands of CMF, non-CMF (ϕ < 120°; ϕ > 240°), respectively.

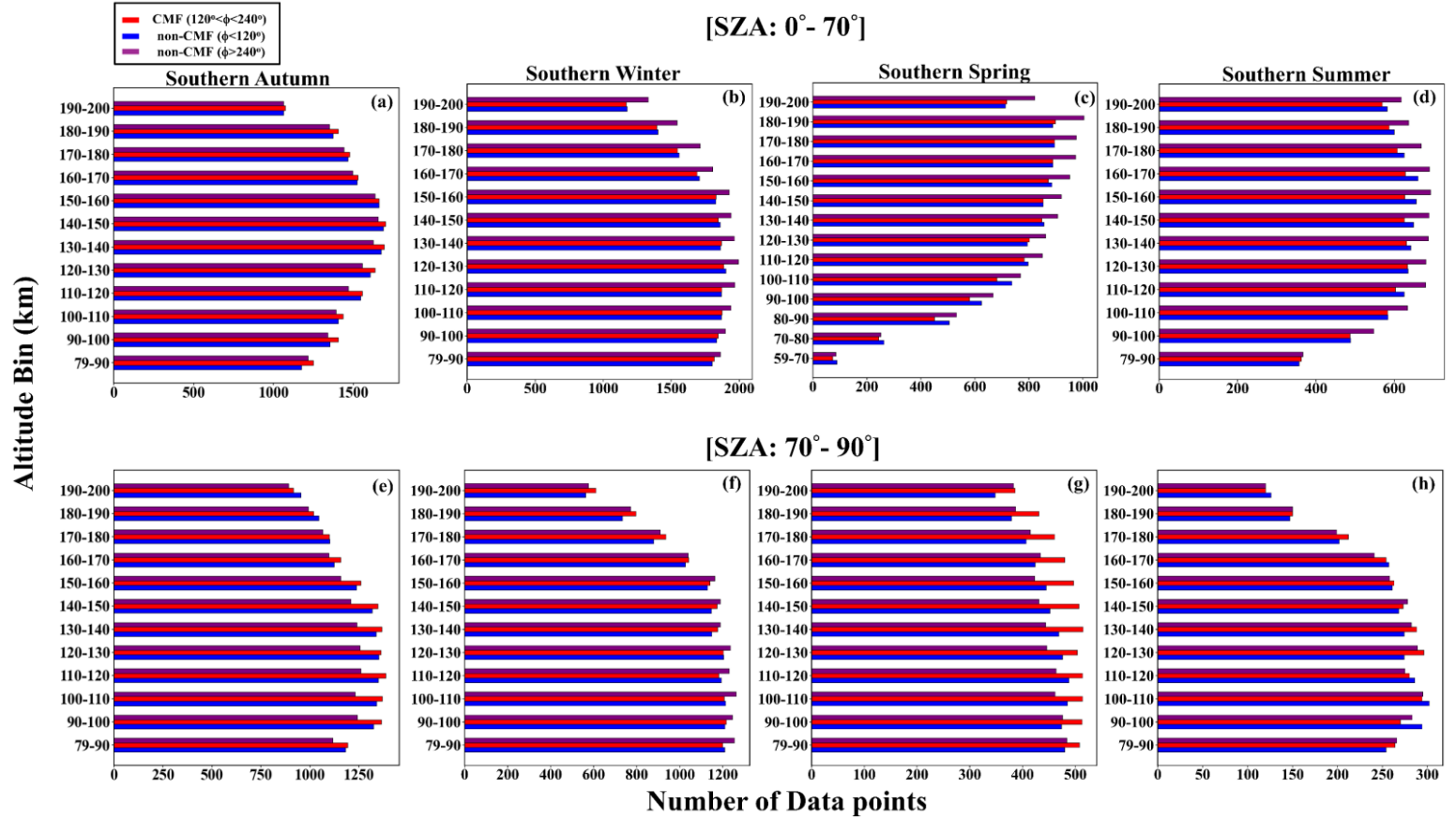

**Figure 1.** Distribution of radiance data points in 10 km altitude bin for different Martian seasons in both dayside and terminator region. The bars plots in red, blue, and purple show the longitudinal bands of CMF, non-CMF (ϕ < 120°), and non-CMF (ϕ > 240°), respectively.

The sub-plots within Figures 2-5 are arranged in the manner that the top panel (a)-(e) shows the median limb radiance profile of CO Cameron band, $CO_2^+$ UVD, $O(^1S)$ 297.2 nm, $O(^3S)$ 130.4 nm and $O(^5S)$ 135.6 nm dayglow emissions in the dayside (SZA: 0°-70°) and bottom panel (f)-(j) represents the variations of dayglow emissions in dayside-terminator region (SZA: 70°-90°). The horizontal error bars are the median absolute deviation.

Figure 2 represents the variation of atmospheric emissions during Southern Autumn season (Ls: 0°-90°) over CMF and non-CMF longitude regions in both the dayside and dayside-terminator regions. It is evident from the figure that all the dayglow emissions in CMF (red), non-CMF (ϕ < 120°; blue) and non-CMF (ϕ > 240°; purple) lie within the median absolute deviations in both dayside (SZA < 70°) and dayside-terminator region (SZA: 70°-90°). The peak altitude of CO Cameron band emission remains around 120 km in both the dayside and terminator regions over crustal and non-crustal longitudes. However, we observe enhanced brightness at peak and below peak of CO Cameron band in the non-CMF ($\phi > 240°$; purple) compared to CMF and non-CMF ($\phi < 120°$; blue) in the dayside-terminator region. However, the same variation in brightness is not observed in the dayside. Similar, enhancement in the brightness is observed for the $CO_2^+$ UVD dayglow in the dayside-terminator region with no change observed in the peak altitude irrespective of the SZA or longitudinal region. The 297.2 nm emission does not vary in crustal and non-crustal region; however, the primary emission peak altitude increases to ~90 km in the dayside-terminator region as compared to dayside peak altitude (~80 km). Similar to the CO Cameron band and $CO_2^+$ UVD emissions, the 130.4 nm and 135.6 nm dayglow emissions exhibit no change in their peak altitude, consistently remaining near 130 km. In contrast, the brightness in the non-CMF ($\phi > 240°$) region shows a depletion on both the dayside and dayside-terminator (SZA: 70° – 90°), with the decrease being significantly more pronounced at 130.4 nm dayglow emission. All dayglow emissions exhibit a decrease in the brightness in the terminator region compared to the dayside. This reduction in the peak brightness is particularly significant for the CO Cameron band (~68%; Figures 2a & 2f), $CO_2^+$ UVD (~68%; Figures 2b & 2g), and 297.2 nm emission (~50%; Figures 2c & 2h). In contrast, the brightness reduction is comparatively smaller for the 130.4 nm (~28%; Figures 2d & 2i) and 135.6 nm (~37%; Figures 2e & 2j) emissions.

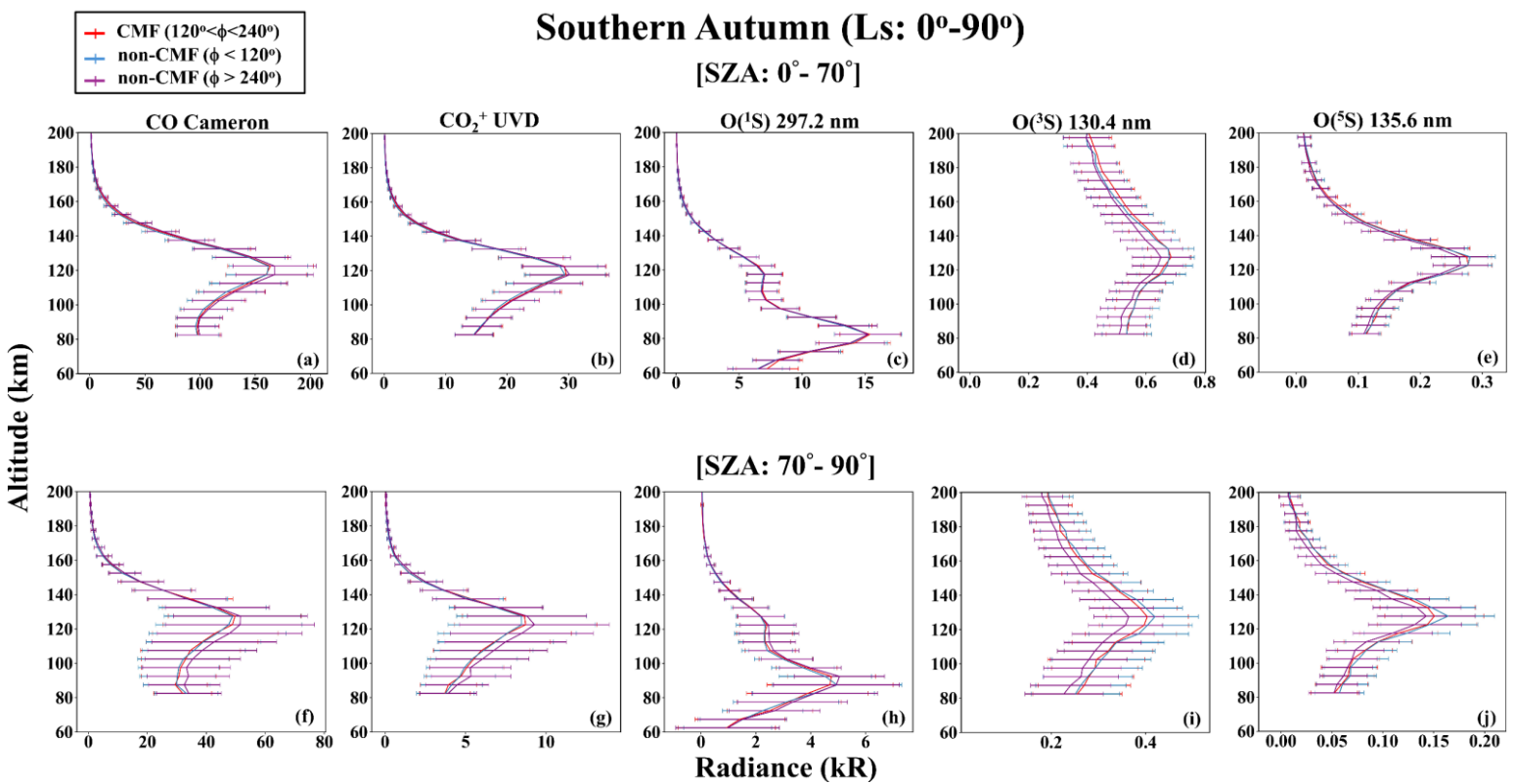

**Figure 2.** Variation of prominent UV dayglow emission in Southern Autumn season (Ls: 0º-90º). Top panel (a)-(e) shows the median radiance profiles in the dayside. Bottom panel (f)-(j) shows the median dayglow profiles in terminator region. The CMF, non-CMF ($\phi < 120°$) and non-CMF ($\phi > 240°$) profiles are shown in red, blue and purple, respectively. The horizontal error bars are median absolute deviation.

Figure 3 depicts the variation of prominent UV dayglow emissions in the Southern Winter season (Ls: 90º-180º). The dayglow emissions limb profiles lie within the median absolute deviation of each other. The CO Cameron band and $CO_2^+$ UVD exhibit no significant variation between crustal and non-crustal regions. Similarly, 297.2 nm dayglow emission shows minimal variation in dayside profiles; however, in the dayside-terminator region the primary emission peak of CMF limb profile displays slight enhancement. Across all three longitude bands, the primary emission peak is located at a higher altitude over the dayside-terminator region (~90 km) compared to the dayside (~80 km). On the contrary, the altitude for CO Cameron band and $CO_2^+$ UVD remain stable at ~120 km in terminator region and dayside. The 130.4 nm and 135.6 nm dayglow emissions also show a stable peak altitude of ~130 km in both dayside and terminator region. During autumn season, the non-CMF ($\phi > 240°$) profile of 130.4 nm dayglow emission brightness exhibits depletion, while the CMF profile demonstrates an increase in brightness over the terminator region. However, 135.6 nm emission does not show any significant variation with only sight enhancement in brightness of CMF profile above the emission peak over the terminator region. For all the dayglow emissions, a lesser magnitude of brightness is observed in dayside-terminator region compared to the dayside. This decrease in the peak brightness is much more prominent for CO Cameron band (Figures 3a & 3f), $CO_2^+$ UVD (Figures 3b & 3g) and 297.2 nm (Figures 3c & 3h) with ~65%, ~70% and ~65%, respectively. On the other hand, the reduction in brightness for 130.4 nm (Figures 3d & 3i) and 135.6 nm (Figures 3e & 3j) are nearly 42% and 33%, respectively.

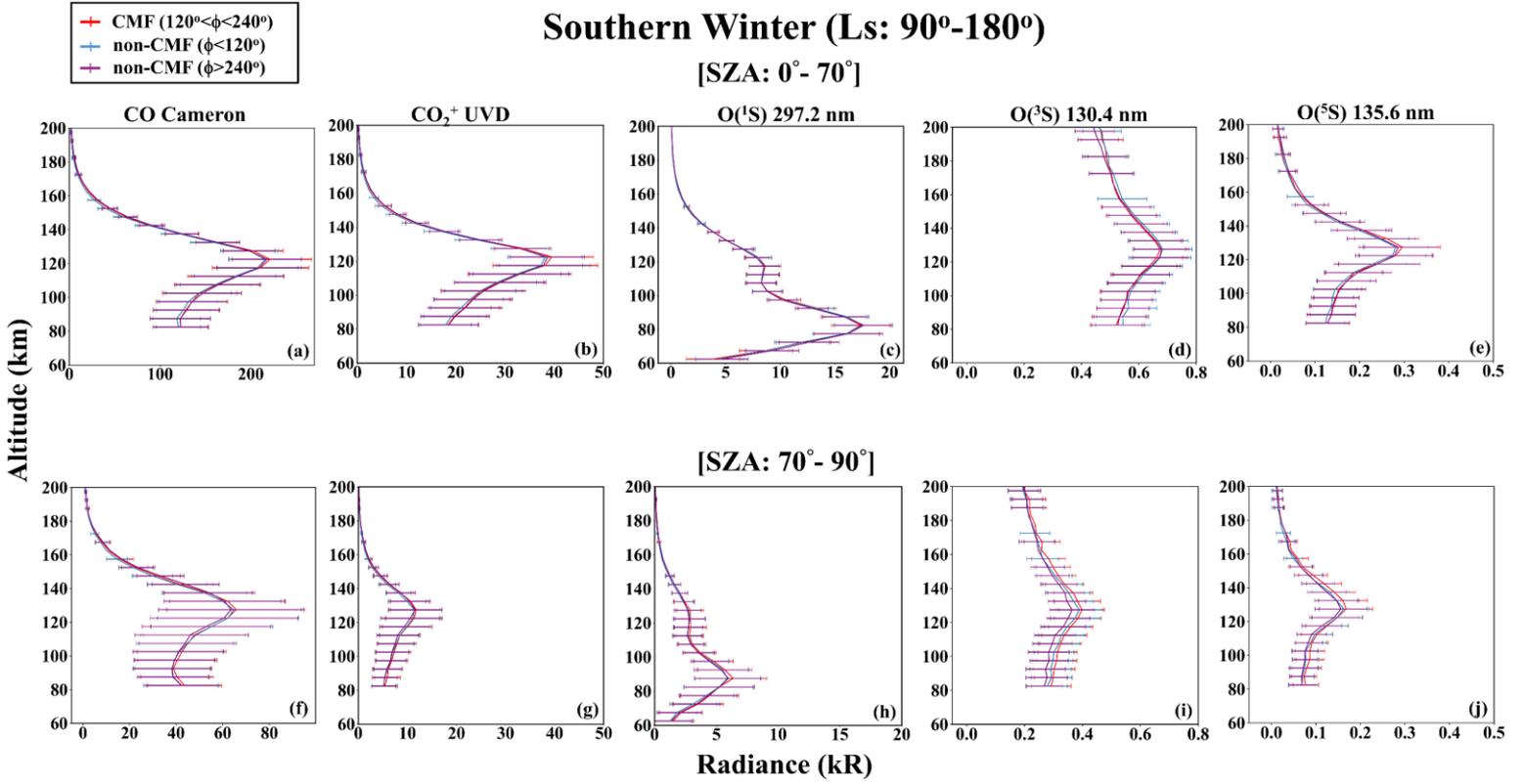

**Figure 3.** Variation of prominent UV dayglow emission in Southern Winter season (Ls: 90°-180°). Top panel (a)-(e) shows the median dayglow profiles in dayside. Bottom panel (f)-(j) shows the median dayglow profiles in terminator region. The CMF, non-CMF ($\phi$ < 120°) and non-CMF ($\phi$ > 240°) profiles are shown in red, blue and purple, respectively. The horizontal error bars are median absolute deviation.

In Figure 4, the variation of median limb brightness in crustal and non-crustal longitudes are shown for Southern Spring season (Ls: 180°-270°). The CO Cameron band (Figure 4a & 4f) and $CO_2^+$ UVD (Figure 4b & 4g) limb profiles have peak altitude of around 140 km of altitude in both dayside and terminator region. The primary emission peak of 297.2 nm also does not show any variation (Figure 4c & 4h) and remains around 100 km of altitude. The emission peak altitudes of 130.4 nm (Figure 4d & 4i) and 135.6 nm (Figure 4e & 4j) also lie near 150 km in the dayside and ~140 km in terminator region. Figures 4f & 4g, showing CO Cameron band and $CO_2^+$ UVD emissions respectively, reveal that in the terminator region, the peak brightness of the non-CMF ($\phi$ > 240°; purple) profile is noticeably lower compared to the CMF (red) and non-CMF ($\phi$ < 120°; blue) profiles. For the CO Cameron band, an enhancement in the brightness is observed at around 80 km altitude in both the dayside and terminator regions (Figures 4a & 4f). Notably, in the dayside-terminator region, this enhancement exceeds the peak brightness. This increase might be due to the artifact from the data reduction or edge effects rather than a genuine atmospheric feature. All dayglow emissions exhibit a decrease in the brightness in the terminator region compared to the dayside. This reduction in the peak brightness is particularly significant for the CO Cameron band (~56%; Figures 4a & 4f), $CO_2^+$ UVD (~55%; Figures 4b & 4g), and 297.2 nm emission (~50%; Figures 4c & 4h). In contrast, the brightness reduction is comparatively smaller for the 130.4 nm (~28%; Figures 4d & 4i) and 135.6 nm (~37%; Figures 4e & 4j) emissions.

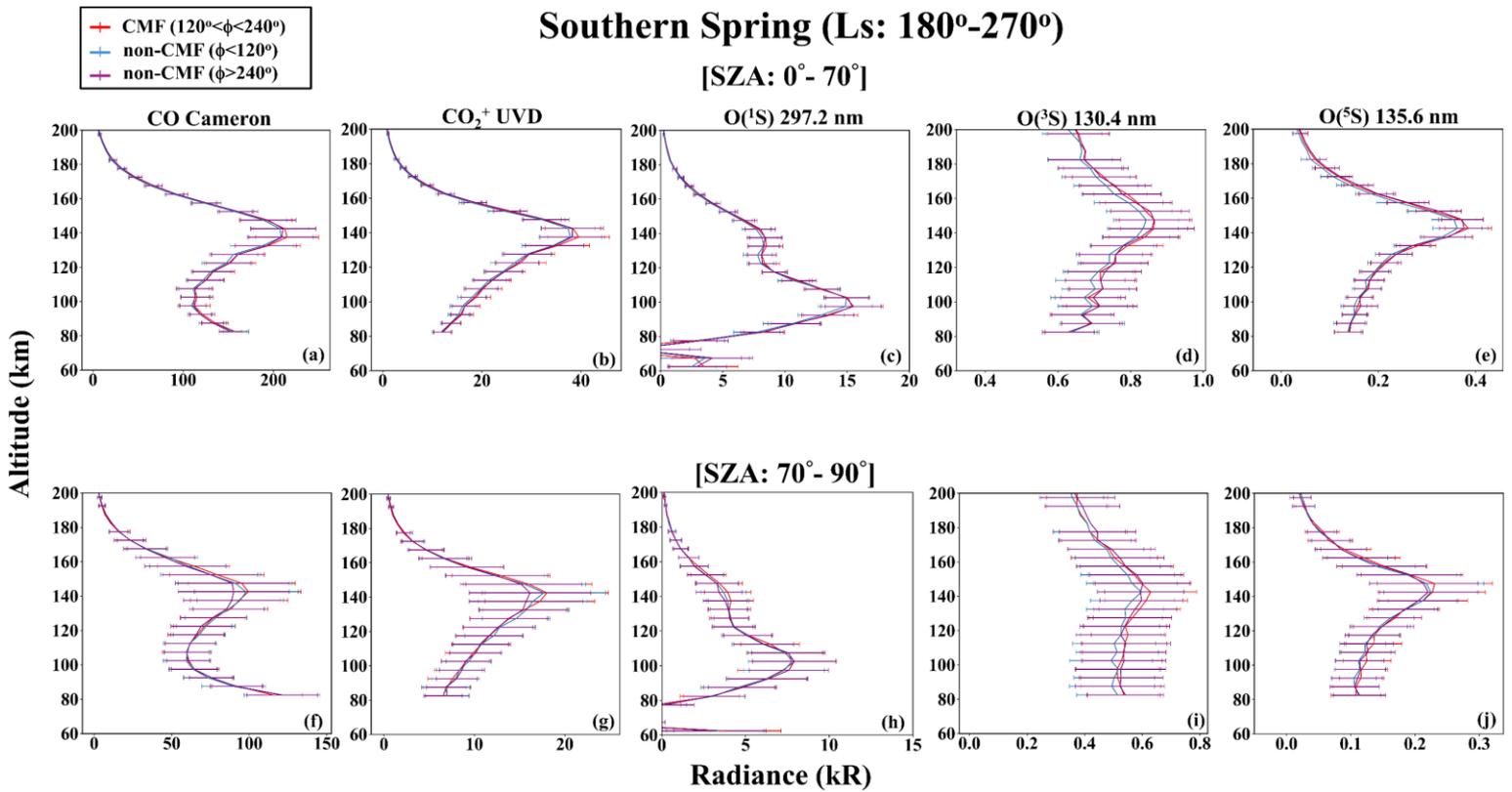

**Figure 4.** Variation of prominent UV dayglow emission in Southern Spring season (Ls: 180°-270°). Top panel (a)-(e) shows the median dayglow profiles in dayside. Bottom panel (f)-(j) shows the median dayglow profiles in terminator region. The CMF, non-CMF (ϕ < 120°) and non-CMF (ϕ >240°) profiles are shown in red, blue and purple, respectively. The horizontal error bars are median absolute deviation.

In Figure 5, the median limb brightness profiles across the crustal and non-crustal longitudes are presented for the Southern Summer season (Ls: 270°–360°). The CO Cameron band (Figures 5a & 5f) and $CO_2^+$ UVD (Figures 5b & 5g) show emission peak altitudes near at 140 km in both the dayside and terminator regions. The primary emission peak for the 297.2 nm emission line also remains constant at around 100 km, without any significant variation between regions (Figures 5c & 5h). Similarly, the 130.4 nm (Figures 5d & 5i) and 135.6 nm (Figures 5e & 5j) emissions exhibit peak altitudes near at 150 km on the dayside and approximately at 140 km in the terminator region. All dayglow emissions exhibit a general reduction in the brightness in the terminator region compared to the dayside. This decrease is consistent across all observed dayglow emissions. The CO Cameron band (~64%; Figures 5a and 5f), $CO_2^+$ UVD (~66%; Figures 5b & 5g), and 297.2 nm emissions (~62%; Figures 5c and 5h) show similar decrease. Similarly, the 130.4 nm (~63%; Figures 5d and 5i) and 135.6 nm (~56%; Figures 5e and 5j) dayglow emissions also display comparable reductions. Overall, the results indicate a similar decrease in the emission intensities from the dayside to the terminator (SZA: 70°-90°) region.

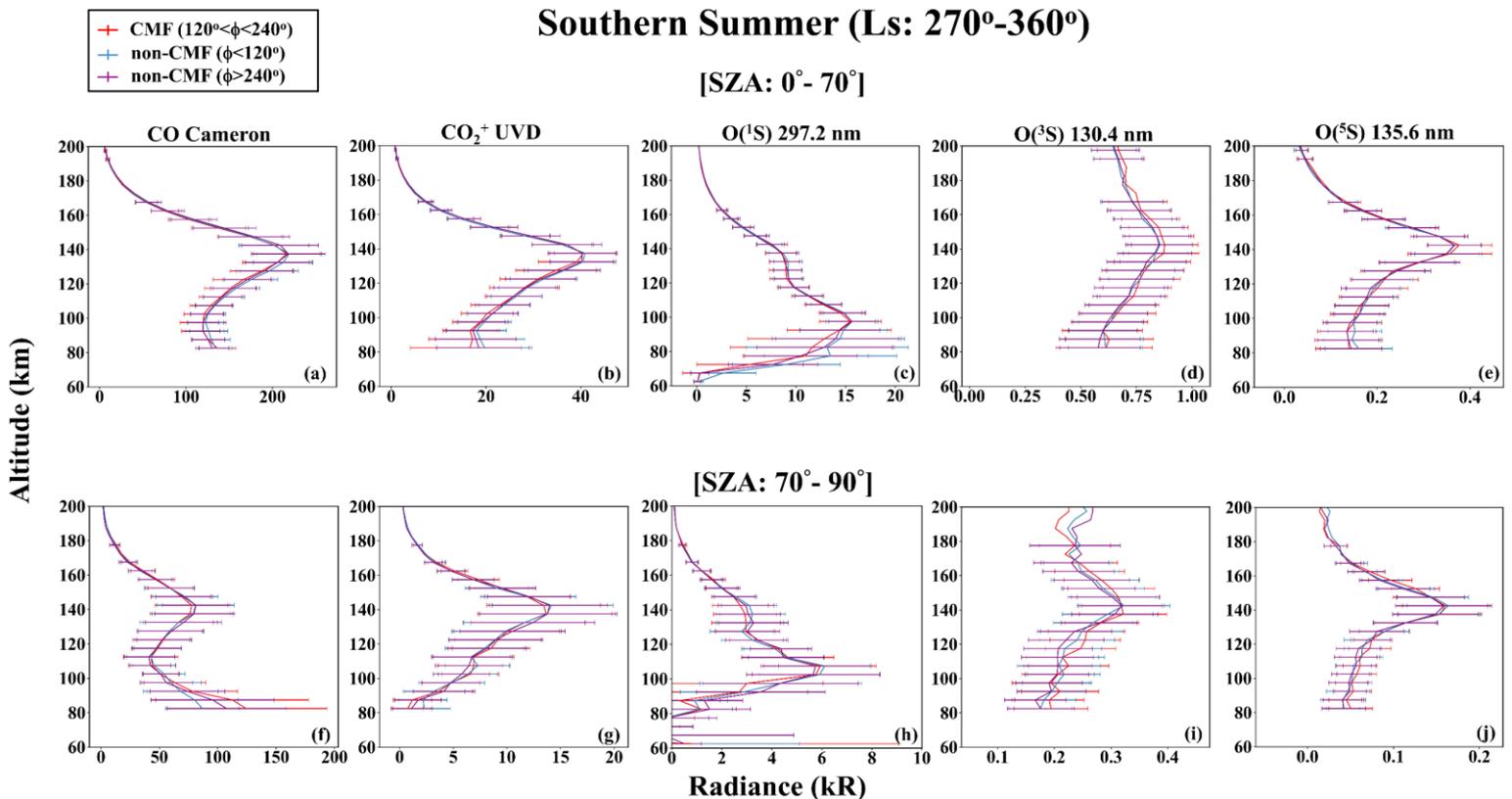

**Figure 5.** Variation of prominent UV dayglow emission in Southern Summer season (Ls: 270°-360°). Top panel (a)-(e) shows the median dayglow profiles in dayside. Bottom panel (f)-(j) shows the median dayglow profiles in terminator region. The CMF, non-CMF (ϕ < 120°) and non-CMF (ϕ < 240°) profiles are shown in red, blue and purple, respectively. The horizontal error bars are median absolute deviation.

## 4. Discussion

The crustal magnetic field in the southern hemisphere of Mars creates a unique environment that significantly influences the plasma and neutral constituents of the planet's upper atmosphere. While prior studies have extensively documented the effects of these fields on the ionosphere above 200 km (Withers et al., 2019; Andrews et al., 2023; Nayak et al., 2024), their influence on the lower atmosphere (below 200 km) remains a subject of ongoing debate (Leblanc et al., 2006; Stiepen et al., 2015; Cui et al., 2018; Chen et al., 2024). Therefore, this investigation, for the first time, provides a comprehensive, multi-seasonal and multi-emission analysis of crustal field effects on Martian dayglow below 200 km altitude. Our study primarily focuses on the variation of prominent Martian dayglow emissions, i.e., CO Cameron band, $CO_2^+$ UVD, $O(^1S)$ 297.2 nm, $O(^3S)$ 130.4 nm, and $O(^5S)$ 135.6 nm in the dayside (SZA < 70º) and dayside-terminator (SZA: 70º – 90º) regions. These dayglow emissions are predominantly produced via the photochemical and electron-impacts processes (Fox and Dalgarno, 1979; Gérard et al., 2019; Jain et al., 2015; Ritter et al., 2019). The behavior of these emissions, and by inferring the possible effects of CMF on dayglow emissions, across different Martian seasons and magnetic regions is examined in detail. In the following sections, we have discussed about the variation of dayglow emissions w.r.t the southern seasons and SZAs.

**4.1 Seasonal Variation**

Our results reveal significant seasonal variations in the peak altitudes and brightness of the dayglow emissions in both CMF and non-CMF regions. For the CO Cameron band and $CO_2^+$ UVD, the peak altitude shifted from approximately 120 km in the Southern Autumn (Figures 2a, 2b, 2e & 2f) and Winter (Figures 3a, 3b, 3e & 3f) to about 140 km in the Southern Spring (Figures 4a, 4b, 4e & 4f) and Summer (Figures 5a, 5b, 5e & 5f). This seasonal shift in emission peak altitudes is consistent with the variability reported in previous MAVEN/IUVS observations, which attributed similar changes to variations in atmospheric density and temperature profiles with Martian seasons (Jain et al., 2015; Gérard et al., 2019). Similarly, the peak altitudes of the atomic oxygen dayglow emissions (297.2 nm, 130.4 nm and 135.6 nm) shifted higher with the seasons. The peak altitude of the dayglow emissions considered in this study are highly dependent on the $CO_2$ density variation and not solely on the solar radiation, but rather on the seasonal changes in Martian atmospheric $CO_2$ content (Gérard et al., 2019; Ritter et al., 2019; Gkouvelis et al., 2020). Therefore, the brightness and altitude of these emissions are expected to vary with seasonal modulation in the upper and middle atmospheric scale height and temperature (Jain et al., 2015, 2023; Bougher et al., 2017).

Brightness variations across seasons were also evident, with emissions generally 30–50% brighter during Southern Spring and Summer compared to Autumn and Winter, correlating with Mars' heliocentric distance (Chaufray et al., 2015). This seasonal brightness enhancement appears largely independent of crustal magnetic field presence, as CMF and non-CMF regions showed remarkably similar seasonal patterns. The minimal differences (<10%) between crustal and non-crustal median profiles of all the emissions suggest that seasonal solar forcing dominates over magnetic field effects for these emissions in the dayside. Furthermore, we have found that all the dayglow emissions exhibited another distinctive feature: peak altitudes remained relatively stable between dayside and dayside-terminator regions in the respective seasons (Figures 2-5). This seasonal stability suggests that during perihelion, the extended thermosphere maintains similar scale heights across a wide range of SZA values (Bougher et al., 2017).

## 4.2 Dayside and Dayside-terminator Variations
### 4.2.1 Dayside

On the dayside (SZA: 0°–70°), emission profiles showed minimal differences between CMF and non-CMF ($\phi$ <120°; $\phi$ >240°) regions across all seasons. The CO Cameron bands and $CO_2^+$ UVD show maximum brightness consistently at similar altitudes, nearly 120 km in the Southern Autumn and Winter, whereas it is around 140 km in the Southern Spring and Summer. These dayglow emissions, primarily driven by photon and electron impact processes during daytime, follow the photo-chemical process proposed by Chapman (1931), where the peak brightness is proportional to the solar EUV flux (Jain et al., 2015). The similar behavior of the CO Cameron band and $CO_2^+$ UVD dayglow emission profiles in the CMF and non-CMF regions (Figures 2-5) suggest that dayside photoelectron production and transport are not significantly modified by crustal fields at smaller SZAs.

For $O(^1S)$ dayglow 297.2 nm emission, the primary emission peak varies around 80-100 km and secondary emission peak at ~120-140 km across both the CMF and non-CMF ($\phi$ <120°; $\phi$ >240°) regions. Previous studies have reported multiple production mechanisms for 297.2 nm dayglow emission, where the primary emissions peak is driven by $CO_2$ photodissociation via solar Lyman-α, and the upper emission peak is dependent on the electron impact process. The observed insignificant changes in the median profiles of 297.2 nm dayglow emission across both the CMF and non-CMF ($\phi$ <120°; $\phi$ >240°) regions (Figures 2-5), signify a key role of photochemical processes irrespective of the crustal magnetic field regions. These outcomes are consistent with the previous studies conducted for plasma and neutral species in the thermospheric region (Flynn et al., 2017; Andrews et al., 2023; Withers et al., 2019; Nayak et al., 2024).

In addition, the peak brightness of $O(^3S)$ 130.4 nm and $O(^5S)$ 135.6 nm dayglow emissions lie at a slightly higher altitudes (~130-150 km) than molecular emissions and remain stable across CMF and non-CMF ($\phi$ <120°), and non-CMF ($\phi$ >240°) regions. This consistency aligns with their production mechanisms: 130.4 nm emission is significantly influenced by the resonant scattering of solar photons, while 135.6 nm emission is primarily produced via excitation caused due to electron impact on atomic oxygen (Chaufray et al., 2015, Ritter et al., 2019).

In totality, an insignificant variation in all dayglow emission profiles suggests a minimal contribution of the crustal fields. This indicates a dominant role of photochemical processes or atmospheric dynamics in the thermospheric region (<200 km). At these altitudes, the collisional frequency is larger than the gyrofrequency, which effectively demagnetized ions and electrons, reducing the crustal magnetic field's impact on plasma dynamics (Akbari et al., 2019; Opgenoorth et al., 2010). In contrast to the dayside, the Martian dayside-terminator region is not solely influenced by the photochemical processes (in the dayside region) but also driven by the transport processes (Cravens et al., 2017; Ram et al., 2024a), which is discussed in the following subsection.

**4.2.2 Dayside-terminator**

In the dayside-terminator region, the peak brightness of CO Cameron band, $CO_2^+$ UVD, and 297.2 nm emissions exhibit larger decrement (50–70%), whereas 130.4 nm and 135.6 nm show a reduction of nearly 28–63% as compared to the dayside. This behavior is consistent irrespective of the CMF and non-CMF regions. The decrease in the brightness from day-to-terminator region aligns with the Chapman theory, in which emission intensity decreases approximately as a function of 'Cosine of SZA' (Jain et al., 2015; Gérard et al., 2019). The smaller percentage deviation in the brightness of 130.4 nm and 135.6 nm emissions could be possible due to the lower loss of atomic O via $CO_2^+$ (Fox, 2004) as compared to the faster dissociative recombination (~1,000 times) for molecular $CO_2^+$ (Fox et al.,2021). Also, the non-uniform reduction across different emissions suggests that the secondary production mechanisms retain relative strength under low solar input at terminator region.

Although, our results show an insignificant change between the CMF and non-CMF ($\phi$ < 120°) regions, we observed enhancements in CO Cameron band and $CO_2^+$ UVD brightness over non-CMF ($\phi$ >240°) region during autumn (purple profile; Figures 2a & 2f). This is possible due to the positive correlation of both CO Cameron and $CO_2^+$ UVD emissions with $CO_2$ density (Jain et al., 2015; Gérard et al., 2019), that leads to the increase in their brightness. On the contrary, 130.4 nm (purple profile; Figures 2d & 2i) and 135.6 nm (purple profile; Figures 2e & 2j) dayglow emissions showed a decrement in their brightness. This is attributed to the strong absorption of emissions (130.4 nm and 135.6 nm) by $CO_2$. Furthermore, $CO_2$ attenuates both the incoming solar radiation and photoelectrons, thereby reducing excitation rates, and absorbs emitted photons along the line of sight (Chaufray et al., 2015; Ritter et al., 2019). Therefore, the increased density of $CO_2$ in non-CMF ($\phi$ >240°) region compared to CMF and non-CMF ($\phi$ < 120°) might be a possible reason for such unusual behavior.

### 4.3 Mechanisms of Masking CMF Effects

The limited differences observed between CMF and non-CMF dayglow profiles, despite well-documented effects of crustal magnetic field on the Martian ionosphere and neutral atmosphere, suggest that multiple compensating mechanisms may obscure magnetic influences on the various emissions.

First, the role of energy-dependent electron shielding. The study by Hara et al. (2018) have shown that in the CMF region horizontal crustal fields preferentially deflect electrons in the energy range of 50–200 eV that are the sufficient energies to produce CO Cameron bands (peak cross-section ~80 eV; Jain et al., 2013). The shielding effect reduces the flux of electrons available for excitation, resulting in a near-zero net effect in the brightness. Furthermore, Chen et al., (2024) found the depletion in the ionospheric peak density in strong crustal magnetic field region due to the inaccessibility of precipitating solar wind particles to the main ionospheric peak altitude, resulting in the weakening of electron impact processes. However, the depletion caused in the ionospheric main peak in crustal field region does not seem to be affecting the dayglow emissions produced via electron impact processes (CO Cameron band and 135.6 nm) in both dayside and terminator region.

Second, the role of thermospheric composition and temperature can play a role. Cui et al. (2018) reported that the strong crustal fields are associated with neutral temperature enhancements of 20–40 K at altitudes of 150–170 km. Such heating can cause atmospheric expansion and lower neutral densities near peak brightness region, counteracting any potential gains due to reduced quenching. The balance between thermally driven scale height increases and density-dependent emission production may thus suppress the observable differences in brightness between crustal and non-crustal regions.

### 5. Summary and Conclusions

This study provides a comprehensive investigation into the influence of Martian crustal magnetic fields (CMF) on dayglow emissions below 200 km in the southern hemisphere, using five Martian years of limb observations from MAVEN/IUVS. By analyzing five key dayglow emissions—CO Cameron bands, $CO_2^+$ UVD, and atomic oxygen lines at 297.2 nm, 130.4 nm, and 135.6 nm—across different seasons and regions of magnetic field strength, the research offers new insights into CMF-dayglow emission interactions at lower altitudes.

The analyses reveal strong seasonal variations in the emission brightness and peak altitudes, closely tied to the changes in the atmospheric density and Mars heliocentric distance. The dayglow emissions peak brightness is larger in magnitude during Southern Spring and Summer compared to Southern Autumn and Winter, consistent with the increased atmospheric scale heights and thermospheric expansion during perihelion. Solar forcing drives altitude shifts in emission peaks (120 km in the Southern Autumn and Winter to 140 km in the Spring and Summer), with molecular emissions (CO Cameron bands, $CO_2^+$ UVD) showing the largest variations due to $CO_2$ density gradients. These seasonal trends appear largely independent of CMF strength, with similar behaviors observed across CMF and non-CMF regions.

Furthermore, the dayglow emission profiles over regions of CMF, non-CMF ($\phi$ <120°; $\phi$>240°) regions display negligible differences. This suggests that, on the dayside, photochemical processes dominate at lower altitudes (below ~200 km), where the influence of magnetic fields is minimal. Additionally, at this altitude the ions and electrons are effectively demagnetized, thus reducing the impact of CMF on dayglow emissions. In contrast, the dayside-terminator region (SZA: 70°–90°) exhibits more variability, including emission enhancements or depletions that vary by emission type, region, and season. Notably, in southern autumn, dayglow emissions like the CO Cameron band, $CO_2^+$ UVD, 130.4 nm and 135.6 nm dayglow emissions show variations in brightness between non-CMF ($\phi$ >240°) and CMF/non-CMF ($\phi$ < 120°) regions (shown in Figure 2). The plausible reason for this behavior of dayglow emissions in the dayside-terminator region could be the enhancement in the $CO_2$ density.

Despite these terminator-region differences, the overall weak or inconsistent influence of CMF observed across most emissions is likely due to various masking mechanisms: (*i*) Electron shielding in CMF regions reduces the flux of <100 eV electrons responsible $CO_2$ dissociation, but the net effect on the brightness of CO Cameron band and 135.6 nm dayglow emissions remains minimal. (*ii*) Furthermore, the thermosphere expands due to CMF-driven temperature increases (~20–40 K) reducing the neutral densities at emission peaks and offsetting potential brightness enhancement.

These comprehensive analyses reveal that Martian crustal magnetic fields minimally influence dayglow emissions below 200 km due to masking by thermal-density compensation, energy-dependent electron shielding. These findings highlight the complex neutral-plasma interactions that suppress possible crustal magnetic field effects. Future missions with in-situ measurements below 200 km and energy-resolved electron mapping are critical to quantify these effects, advancing our understanding of Martian atmospheric evolution and impact of space weather.

**6. Data availability**

MAVEN datasets of IUVS Level 1c, version_13, revision_01 utilized during this work are available through the NASA Planetary Data System (PDS) (https://atmos.nmsu.edu/PDS/data/PDS4/MAVEN/iuvs_processed_bundle/l1c/limb/). The MAVEN/IUVS derived data products utilized and produced during this work can be found in the Zenodo repository (Sharma et al., 2025b).

## 7. Acknowledgements

A. R. Sharma and L. Ram acknowledge the fellowship from the Ministry of Education, Government of India for carrying out this research work. We sincerely acknowledge the NASA PDS, MAVEN teams, especially IUVS team members for the dayglow dataset. This work is supported by the Ministry of Education, Government of India.